\documentclass[prb,twocolumn,superscriptaddress]{revtex4-2}
\usepackage[utf8]{inputenc}
\usepackage{amsmath}
\usepackage{physics}
\usepackage{graphicx}
\usepackage{tikz}
\usetikzlibrary{patterns,snakes}
\usepackage{amsfonts}   
\usepackage{comment}

\DeclareMathOperator{\arccosh}{arccosh}
\newcommand{\nobarfrac}{\genfrac{}{}{0pt}{}}

\definecolor{taylorswift}{rgb}{0.0862745098,0.4666666667,0.3411764706}
\definecolor{fearless}{rgb}{0.8862745098,0.6117647059,0.2823529412}
\definecolor{speaknow}{rgb}{0.4588235294,0.2274509804,0.4980392157}
\definecolor{red}{rgb}{0.6509803922,0.1254901961,0.2705882353}
\definecolor{TS1989}{rgb}{0.1803921569,0.6,0.9764705882}
\definecolor{reputation}{rgb}{0.1450980392,0.1490196078,0.1529411765}
\definecolor{lover}{rgb}{0.8392156863,0.2117647059,0.5529411765}

\begin{document}

\title{Emergence of quasiperiodic behavior in transport and hybridization properties of clean lattice systems}

\author{Cecilie Glittum}
\author{Antonio Štrkalj}
\author{Claudio Castelnovo}
\affiliation{T.C.M. Group, Cavendish Laboratory, JJ Thomson Avenue, Cambridge CB3 0HE, United Kingdom}

\begin{abstract}
    Quasiperiodic behaviour is mostly known to occur in systems with enforced quasiperiodicity or randomness, in either the lattice structure or the potential, as well as in periodically driven systems. Here, we present instead a rarer setting where quasiperiodic behaviour emerges in clean, non-driven lattice systems. We illustrate this through two examples of experimental relevance, namely an infinite tight-binding chain with a gated segment, and a hopping particle coupled to static Ising degrees of freedom. We show how the quasiperiodic behaviour manifests in the number of states that are localised by the geometry of the system, with corresponding effects on transport and hybridisation properties. 
\end{abstract}
\date{\today}

\maketitle
%
%

\section{\label{sec:intro}
Introduction}
Some of the most striking discoveries in physics occur when simple systems are found to exhibit unexpectedly complex behaviour. 
These are often prime examples of the beauty and elegance of mathematical modelling that succeeds in describing the world around us. 
Examples include the chaotic motion of the double pendulum~\cite{Landau1960,Shinbrot1992} and the quantum Hall effect of a 2D electron gas in a strong applied field~\cite{Ezawa2013}.

Without claim of drawing a comparison in importance, we present here a simple result that fits well in this category. In one of its simplest forms, our finding shows that the number of states localised~\footnote{The reader should bear in mind here, for reference, the concept of Wannier-Stark localisation~\cite{Bloch1929,Zener1934,Wannier1962}.} on a gated portion of an infinite tight-binding chain forms a \emph{quasiperiodic} sequence in the number of gated sites. Equivalently, the number of resonance peaks in the conduction along the chain also forms a quasiperiodic sequence. This happens for both negative and positive gate potential, within the bounds of the tight-binding band. 
We provide both an exact solution to the problem as well as a simple, elegant and intuitive (albeit only effective) derivation of the same solution. 

Our results are of direct relevance to several experimental settings -- thinking for example about cold atoms~\cite{Krinner2015, Lebrat2018}, trapped ions~\cite{Monz2011,Friis2018}, quantum dot arrays~\cite{Zajac2016}, superconducting cubits~\cite{Guo2021} and other quantum simulators~\cite{Altman2021} -- and can be readily verified in a laboratory. 
Moreover, in addition to this simplest formulation of the problem, we show that a similar phenomenology is expected to play a role in other settings, for instance when particles hopping on a lattice are coupled to underlying spin degrees of freedom, as in the Falicov-Kimball model~\cite{Falicov1969} and in systems that exhibit disorder-free localisation~\cite{Smith2017}. 

Quasiperiodic systems have received significant attention of late (see for instance Refs.~\onlinecite{Schreiber2015,Bordia2017,Varma2019,Mace2019,Johnstone2019,Ghadimi2020,Sbroscia2020,Flicker2020,Szabo2020,Strkalj2021,Oktel2021,Padavic2021,Friedman2022,Agrawal2022,Strkalj2022}). In most cases however the quasiperiodicity is built into the system, e.g., via quasiperiodic modulation of on-site potentials and hopping amplitudes, or induced by incommensurate periodic driving; then, properties deriving from it are studied. Examples where quasiperiodicity is not seeded but rather emerges from periodic constituents are rare (most notably this was uncovered to result from the interplay of lattice filling fractions and interactions in Refs.~\onlinecite{Gopalakrishnan2013,Sagi2016}; however, see also Ref.~\onlinecite{Mivehvar2019} for an example of a quasicrystalline potential emerging from collective light scattering). In our setting, the quasiperiodic behaviour emerges in a clean system made of non-driven periodic components -- something that does not happen often in nature. 
We envision that the simple mechanism discussed in this work is likely to operate in interesting ways in other condensed matter systems, and possibly in higher dimensions as well. 
%
%

\section{\label{sec:gated_segment}
Gated chain}
%
\begin{figure}
    \centering
    \includegraphics[scale=1]{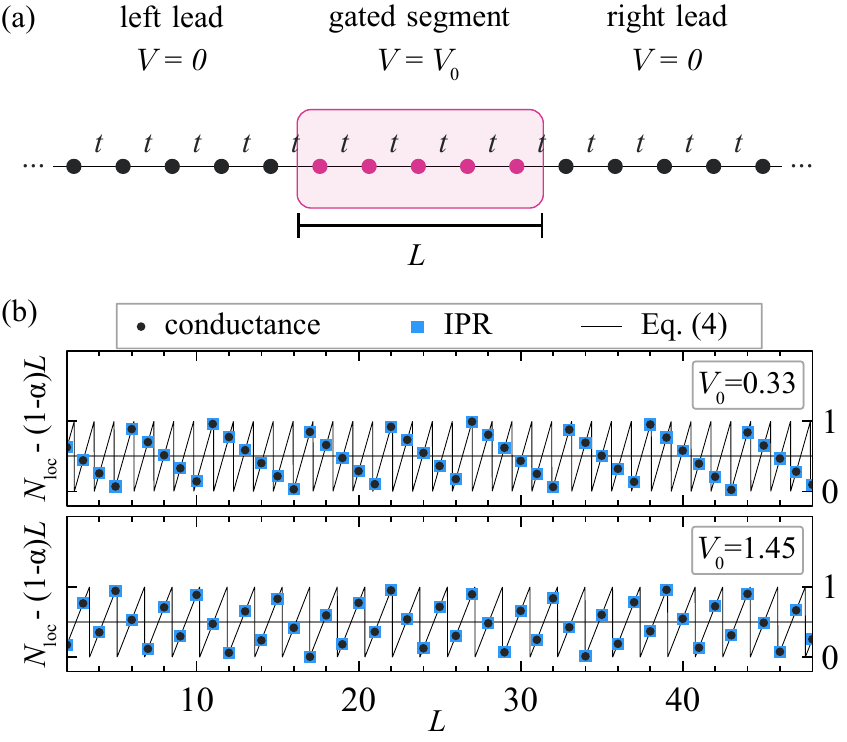}
    \caption{
    (a) Illustration of the gated chain. The solid black dots illustrate the infinite leads, at potential $V=0$. The solid pink dots illustrate the gated segment of length $L$ at potential $V_0$, here for $L=5$.
    (b) Number of states localised in the gated region $N_{\mathrm{loc}}$ as a function of its length $L$. We subtracted the linear slope $(1-\alpha)L$ to emphasise the quasiperiodic nature of the fluctuations following from sampling at integer $L$. Numerically obtained data from conductance simulations (black dots) and IPR (blue squares) overlap and agree perfectly with the analytical prediction in Eq.~\eqref{eq:exact_solution} (black line).  
    }
    \label{fig:system_Nloc}
\end{figure}
\begin{figure*}[ht]
    \centering
    \includegraphics[scale=0.98]{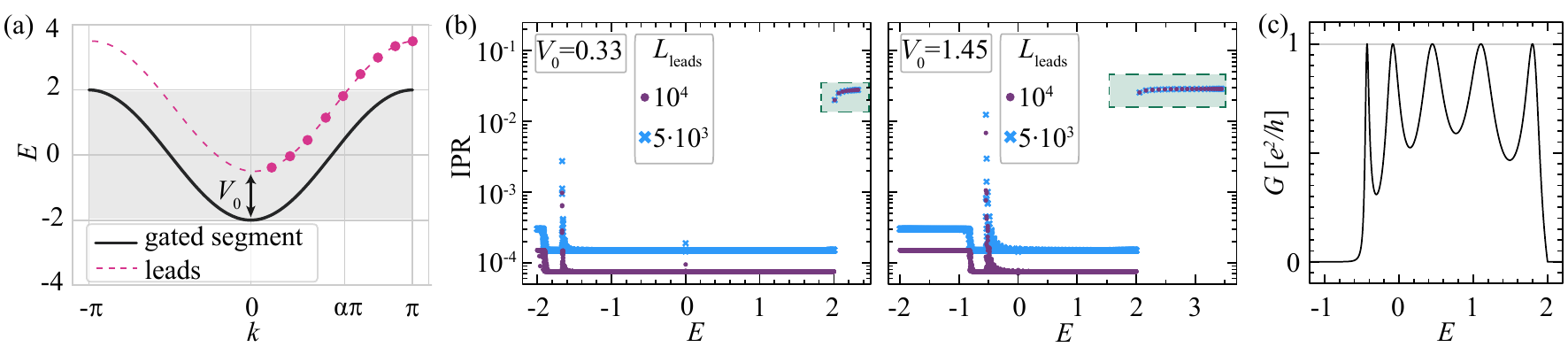}
    \caption{
    (a) Dispersion of the infinite leads (black solid line) and of the gated segment in the $L \rightarrow \infty$ limit (pink dashed line). The two dispersions overlap for $k\in [-\alpha \pi, \alpha \pi]$. We also show the energies of an isolated gated segment with $L=9$ plotted as a function of $q_n = \pi n/L$; $n = 1,2, \ldots, L$ (pink solid dots). Counting the number of discretised energies falling outside of the bandwidth of the leads gives the number of states localised on the gated segment, which in this case is $4$ states.
    (b) IPR for a gated segment of length $L = 50$ with $V_0 = 0.33$ (left) and $V_0=1.45$ (right), comparing two different lead sizes, $L_{\mathrm{leads}} = 5\cdot 10^3$ and $L_{\mathrm{leads}} = 10^4$. States with $E > 2$ (highlighted by a green rectangle) are localised inside the gated segment and their IPR is independent of the lead size. 
    (c) Conductance through the whole system as a function of energy of the incident particle. Here we used $L=9$ and $V_0=1.45$. 
    We remind that potentials and energies are expressed in units of the hopping amplitude $t=1$.
    }
    \label{fig:dispersions_IPR_Nloc_conductance}
\end{figure*}
We study a tight-binding chain of lattice spacing $a=1$ and hopping amplitude $t=1$, with a segment of length $L$ gated at potential, i.e., on-site energy, $V_0$. The rest of the system (i.e., the left and right leads) is kept at the reference potential $V=0$. 
The Hamiltonian of the system is given by
\begin{equation}\label{eq:hamiltonian_system_leads}
    H = -t\sum_{i} \left[c^\dagger_{i+1}c_{i} + \mathrm{h.c.}\right] + \sum_{i} V_i c^\dagger_{i}c_{i},
\end{equation}
where $V_i$ is finite on the gated segment and vanishes on the leads. The system is illustrated in Fig.~\ref{fig:system_Nloc}(a).

Due to the mismatch in on-site potential between the gated segment and the leads, some states localise on the gated segment, while other states are extended over the whole system. 
Our goal is to investigate the nature of these localised states, and more precisely how their number changes with the system parameters $L$ and $V_0$. As we detail below, the number of states Wannier-Stark-localised on the gated segment forms a \emph{quasiperiodic} sequence as a function of the discrete length $L$ of the segment; see Fig.~\ref{fig:system_Nloc}(b). 

By choosing an appropriate ansatz for the wave function, we look for eigenstates of the Schr\"{o}dinger equation that are exponentially decaying on the leads (i.e., localised on the gated segment); 
see App.~\ref{app:Analytical N_L} for details and App.~\ref{app:WFProfiles} for examples of wave function profiles. 
We find that the number of such states corresponds to the number of solutions of either of the following equations:
\begin{eqnarray}
 -e^{\arccosh \left[V_0/2 -  \cos k\right]} &=& \frac{ \cos\left[k - k\frac{L+1}{2} \right]}{ \cos \left[ - k\frac{L+1}{2} \right]}, \label{eq:gated-k-eq-1}
\\ 
 -e^{\arccosh \left[V_0/2 -  \cos k\right]} &=& \frac{ \sin \left[k - k\frac{L+1}{2} \right]}{ \sin \left[ - k\frac{L+1}{2} \right]}
\, . \label{eq:gated-k-eq-2}
\end{eqnarray}
The task can be conveniently re-cast into counting the number of $q_n = n \pi/L $ with $n\in \{1, \dots, L\}$ larger than $\alpha\pi \equiv \arccos{\left(\frac{V_0}{2} - 1\right)}$, which is the point in the Brillouin zone where the dispersion of the isolated infinite gated segment intersects the band edge of the isolated leads; see Fig.~\ref{fig:dispersions_IPR_Nloc_conductance}(a). 
Knowing $q_n$ and $\alpha$, one can show that the number of states localised on the gated segment is given by
\begin{equation}\label{eq:exact_solution}
    N_{\rm loc} = L - \mathrm{floor}\left( \alpha L \right)
\, ,
\end{equation}
which is a linearly growing function of $L$ with slope $1-\alpha$, and it exhibits quasiperiodic fluctuations whenever $\alpha$ is irrational. 

We note in passing that the same result can be obtained using Levinson's theorem~\cite{Levinson1949}, appropriately modified for a lattice model, from the phase difference between the states at the bottom and top of the band 
(see App.~\ref{app:Analytical N_L Levinson}). 

The quasiperiodicity can be distinctly seen in Fig.~\ref{fig:system_Nloc}(b), where we also verify the perfect agreement between the analytical result, Eq.~\eqref{eq:exact_solution}, and the numerical calculation of $N_{\rm loc}$ using both the inverse participation ratio (IPR) and transport measurements. Albeit simple, we find this result surprising since the leads and the gated segment are periodic, and there is no additional quasiperiodicity or randomness present in the system. 
%
%

\subsection{Simple counting argument}
It is tempting to interpret the above $q_n$ as the wave vector of a tight-binding chain of length $L$ with open boundary conditions. The argument appears then to be counting the number of discrete eigenenergies of the gated segment, whose dispersion in the limit of infinite length is $E_{\mathrm{gated}}(k) = V_0-2\cos k$, that fall outside the continuous band of the leads, of dispersion $E_{\mathrm{leads}}(k) = -2\cos k$. 

The two dispersions intersect at $k = \pm \alpha\pi$, as illustrated in Fig.~\ref{fig:dispersions_IPR_Nloc_conductance}(a), and one can see that the discrete energies of the gated segment fall outside the band of the leads if $q_n > \alpha\pi$. Recalling that $n = 1,2, \ldots, L$, one can straightforwardly count such states and obtain $L - \mathrm{floor}\left( \alpha L \right)$ for the number of states localised on the gated segment, which coincides with the exact solution, Eq.~\eqref{eq:exact_solution}. 

This is however only an intuitive, and somewhat incorrect, picture which happens coincidentally to give the correct answer. In the exact calculation, the $q_n$'s are simply a tool to count the number of solutions and have no direct physical interpretation. 
%
%

\subsection{Probing localised states}
One way to probe the localised states is by computing the IPR of the eigenstates of the Hamiltonian~\eqref{eq:hamiltonian_system_leads},
\begin{equation}
    \mathrm{IPR}(E_n) = \frac{\sum_{j} |\psi_j(E_n)|^4}{\sum_{j} |\psi_j(E_n)|^2} \, .
\end{equation}
The IPR is known to scale as (i) $\mathcal{O}(1/L_{\mathrm{tot}})$ for states extended over the whole system of length $L_{\mathrm{tot}} = 2 L_{\mathrm{leads}} + L$, where $L_{\mathrm{leads}}$ is the number of sites in a single lead; (ii) $\mathcal{O}(1/(2L_{\mathrm{leads}}))$ for states localised on the leads (when one allows for a small matrix element due to tunnelling across the gated segment); and (iii) $\mathcal{O}(1/L)$ for states localised only on the gated segment.
We use exact diagonalisation and compare the results for two different sizes of leads, $5\cdot 10^3$ and $10^4$ sites, and for different gate potentials. The results for a gated segment with $L = 50$ sites are shown in Fig.~\ref{fig:dispersions_IPR_Nloc_conductance}(b), for $V_0 = 0.33$ and $V_0 = 1.45$ (recall that we work in units where $t=1$). 
We see that all the states with energy $E > 2$ have an IPR that is independent of the size of the leads, meaning that these states are localised on the gated segment. The aforementioned states cannot hybridise with the states belonging to the leads as their energy falls outside their bandwidth. On the other hand, states falling within the bandwidth of the leads hybridise and delocalise. 

The steplike behaviour seen in Fig.~\ref{fig:dispersions_IPR_Nloc_conductance}(b) is a numerical artefact of the diagonalisation algorithm. When the energy of the eigenstates is smaller than $-2+V_0$, the particle cannot propagate across the barrier; however, a small tunnelling amplitude exists for finite values of $L$ such that the eigenstates found by diagonalisation are even and odd superpositions of the states on the left and on the right lead, and the IPR is approximately $1/(2L_{\rm leads})$ (which in the figure is indistinguishable from $1/(2L_{\rm leads}+L)$, the IPR of delocalised states). 
As we continue to lower the energy of the eigenstates however, the tunnelling amplitude is correspondingly suppressed; when it falls below numerical accuracy, the diagonalisation routine is no longer able to find even and odd superpositions of states on the left and on the right lead, but rather sees the left and right lead states as eigenstates of the system. This results in an IPR~$\sim 1/L_{\rm leads}$ and in the step increase observed when we consider the lowest energy eigenstates. 

Another way to directly probe the aforementioned localised states is via energy-dependent transport, i.e., via the conductance 
\begin{equation}
    G(E) = \frac{e^2}{h} \mathcal{T}(E) \mathcal{T}^{*}(E) \, ,
\end{equation}
where $e$ is the charge of the particle, $E$ is its incident energy, and $\mathcal{T}$ is the transmission coefficient. The above conductance is observed to have an oscillatory behavior in energy, with maxima reaching the value $e^2 / h$, and the distance between them determined by the 
length $L$ (see App.~\ref{app:Analytical cond}). 
In Fig.~\ref{fig:dispersions_IPR_Nloc_conductance}(c) we show the numerical result from the simulation of a system consisting of two identical and infinite leads and a gated segment of $L=9$ sites (see App.~\ref{app:DependenceOnV0} for a plot of the conductance as a function of $V_0$ for fixed $L$). 
The peaks in conductance occur once $E$ is in resonance with the energies of the gated segment that hybridise with the leads, i.e., the energies that lie within the bandwidth of the leads. 
By counting the number of peaks $N_{\rm peaks}$, whose maxima reach the value $e^2 / h$, we can extract the number of states that localise on the gated segment as $N_{\rm loc} = L - N_{\rm peaks}$. The result for the gated chain discussed before is shown in Fig.~\ref{fig:system_Nloc}(b). We observe perfect agreement with $N_{\rm loc}$ obtained by IPR, as well as with the analytical prediction~\eqref{eq:exact_solution}. Interestingly, a simple analytical calculation of conductance which assumes that the leads do not affect the dispersion of the gated segment gives the same expression for $N_{\rm loc}$ as in Eq.~\eqref{eq:exact_solution}; 
see App.~\ref{app:Analytical cond}. 


%
%

\section{\label{sec:antiferro_chain}
Ferromagnetic segment in an antiferromagnetic chain}
\begin{figure}
    \centering
    \includegraphics[scale=1]{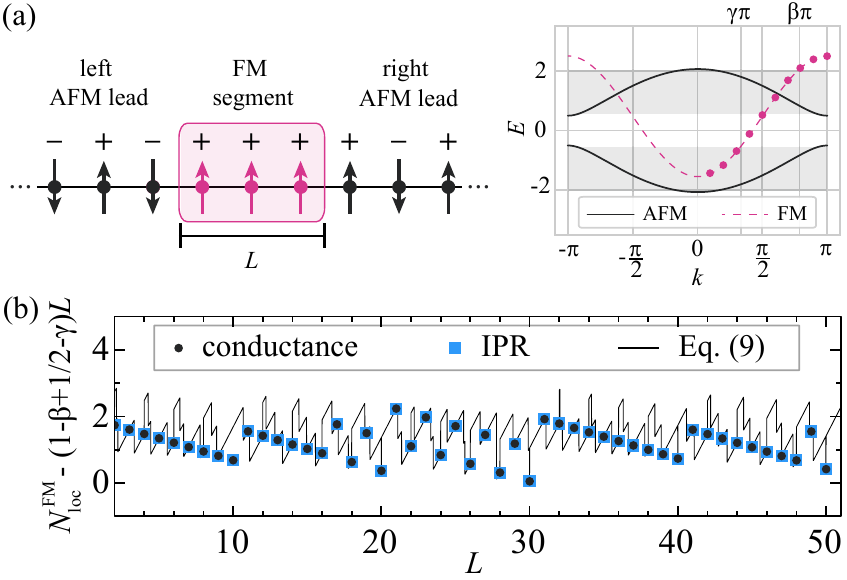}
    \caption{
    (a) Left: Illustration of the FM segment in an AFM chain. Black points and arrows illustrate the infinite AFM leads and pink ones illustrate the FM segment of length $L$, here for $L=3$. Right: The AFM and FM dispersions for $W = 0.5$. The FM band overlaps with the lower AFM band in the region $k \in [-\gamma \pi, \gamma \pi]$ (for $W \leq 1$) and with the upper AFM band in the region $k \in [-\beta \pi,-\pi / 2]$ and $k \in [\pi / 2, \beta \pi]$. 
    (b) Number of states localized in the gated region as a function of its length for $W=0.1$. We subtracted the linear slope $(1 - \beta + 1/2 - \gamma) L$ to emphasise the quasiperiodic nature of the fluctuations following from sampling at integer $L$. We observe a perfect agreement between numerical data obtained from conductance (black dots), IPR (blue squares) and the analytical prediction in Eq.~\eqref{eq:FMinAFM}. 
    }
    \label{fig:AFMandFMdisp_and_Nloc}
\end{figure}

Similar quasiperiodic behaviour occurs also in related systems of relevance to other experimental settings. Consider for instance the case of an Ising chain where $+$ and $-$ correspond to an on-site energy $\pm W$ for a tight-binding particle along the same chain, with hopping amplitude $t=1$ (where without loss of generality we set $W>0$). 
Specifically, take an infinite antiferromagnetic (AFM) chain with a ferromagnetic (FM) insertion $++ \ldots +$ of length $L$ (see Fig.~\ref{fig:AFMandFMdisp_and_Nloc}(a) for a precise illustration of how we define the inserted FM segment). 

The dispersion of the AFM leads is given by
\begin{equation}\label{eq:EAFM}
    E_{\mathrm{AFM}}(k) = \pm \sqrt{W^2 + 2\left( 1 + \cos k \right)},
\end{equation}
which exhibits a gap near zero energy, in contrast with the case considered earlier. 
The dispersion of the FM segment is instead given by the usual relation 
\begin{equation}
    E_{\mathrm{FM}}(k)=W-2\cos k.
\end{equation}
Both dispersions are illustrated in Fig.~\ref{fig:AFMandFMdisp_and_Nloc}(a). Note that the FM segment overlaps with both bands of the leads for $W \leq 1$ and only with the upper band for $W > 1$. The FM band overlaps with the lower AFM band for $k \in [-\gamma\pi, \gamma\pi]$ (for $W \leq 1$), where $\gamma \pi \equiv \arccos (W)$, and with the upper AFM band for $k \in \left\{[-\beta\pi, \pi/2], [\pi/2, \beta\pi]\right\}$ with $ \beta\pi \equiv \arccos \left[\frac{1}{2}\left( W - \sqrt{W^2 + 4} \right)\right]$ (the values of $\beta$ and $\gamma$ are visually represented in the right panel of Fig.~\ref{fig:AFMandFMdisp_and_Nloc}(a) for added clarity).

All states falling outside of the AFM bands correspond to states localised on the FM domain, falling off exponentially when moving into the AFM leads (see App.~\ref{app:WFProfiles} for examples of wave function profiles). 
Using a corresponding ansatz for the lattice wave function, as we did for the gated segment earlier, one can explicitly find the localised eigenstates of the system 
(see App.~\ref{app:Analytical N_L^FM}) 
by solving equations similar to Eqs.~\eqref{eq:gated-k-eq-1} and~\eqref{eq:gated-k-eq-2}.
As shown in App.~\ref{app:Analytical N_L^FM}, 
the number of states localised on the FM segment with energies above the upper AFM band can be found by counting the number of $q_n = n \pi/L$ with $n \in \{ 1 , \dots , L\}$ larger than $\beta \pi$, i.e., $L-\mathrm{floor}(\beta L)$.
Similarly, the number of states localised on the FM segment with energies in the gap of the AFM dispersion can be found by finding the number of $q^{\infty}_n = n \pi/(L+2) < \pi/2$ and if $W\leq 1$ then subtracting the number of $q_n = n \pi/L \geq \gamma \pi$. This leads to $\mathrm{ceil}\left(L/2\right) - \mathrm{floor}\left(\gamma L\right)$ localised states in the gap for $W \leq 1$ and $\mathrm{ceil}\left(L/2\right)$ for $W > 1$.
The total number of states localised on the FM domain is thus given 
by (App.~\ref{app:Analytical N_L^FM})%
~\footnote{A similar simple counting argument can be used here as we did in the case of the gated segment. However, it seems only to be able to reproduce the exact expression in Eq.~\eqref{eq:FMinAFM} up to an additive unit constant.}
\begin{align}\label{eq:FMinAFM}
    N^{\mathrm{FM}}_\mathrm{loc} =& \begin{cases}
    L - \mathrm{floor}\left(\beta L \right) + \mathrm{ceil}\left(L/2\right) - \mathrm{floor}\left(\gamma L\right), & W \leq 1\\
    L - \mathrm{floor}\left(\beta L \right) + \mathrm{ceil}\left(L/2\right), & W > 1
    \, .
    \end{cases}
\end{align}

To confirm the above formula, we numerically extract the number of states localised on the FM domain, $N^{\mathrm{FM}}_{\rm loc}$, using both IPR and conductance. The results for $N^{\mathrm{FM}}_{\rm loc}$ as a function of size of the FM domain $L$ are shown in Fig.~\ref{fig:AFMandFMdisp_and_Nloc} (b), where we observe perfect agreement with the analytical prediction, Eq.~\eqref{eq:FMinAFM}. 

As an aside, we note that it is possible to numerically compute the number of localised states on the FM domain more efficiently by counting the number of sign changes in the Sturm sequence obtained by relating the subdeterminants of the Hamiltonian~\cite{Ortega1960}. This enables one to treat larger system sizes at a relatively low computational cost compared to full diagonalisation. 
%
%

\section{Conclusions}
In this paper, we demonstrated how quasiperiodic behaviour can emerge in clean periodic systems without driving, borne out of a mechanism similar to the one that gives rise to Wannier-Stark localisation~\cite{Bloch1929,Zener1934,Wannier1962}. 
%
%
For instance, we showed that the number of localised states on a gated segment of a tight-binding chain (equivalently, the number of peaks in the conductivity along the chain) is a quasiperiodic function of the length of the segment. A similar behaviour arises when the tight-binding particle is instead subject to on-site coupling to static Ising degrees of freedom, which form an antiferromagnetic chain with a ferromagnetic segment of finite length inserted into it. 
We provide both an exact solution as well as an intuitive effective description of the phenomenon, and we discuss how it is directly relevant to various settings, from gated quantum dot arrays to systems described by the Falicov-Kimball model. Similar behaviour can be expected in general if a finite segment is inserted into spatially translationally invariant leads. Quasiperiodicity should arise when the dispersion of the inserted segment intersects the band edge of the leads at some $k = \alpha \pi$, with $\alpha$ irrational.

(We note in passing that when both gating and coupling to an underlying spin pattern are present, additional features may appear in the form of states localised at the boundaries of the segment.) 

The mechanism discussed in our work is not inherently limited to 1D noninteracting systems and the experimental platforms mentioned in the introduction. Similar mechanisms are expected to operate for example in strongly interacting systems, e.g., quantum wires described by Luttinger liquid theory, that are coupled to leads of finite bandwidth~\cite{Safi1995,Nazarov1997,Gutman2010,Strkalj2019}.
They should also be relevant to two- and higher-dimensional gated systems. One can trivially see it if one uses a separable potential: $V(x,y) = V_0 [\Theta(x) + \Theta(y)]$, where $\Theta$ is a function that takes the value $1$ on a segment of length $L$ and vanishes elsewhere in the system. This potential results in an infinite cross pattern in 2D (of potential $V_0$), with a raised square at its core (of potential $2 V_0$ and size $L \times L$). The exact analytical solution follows directly from the 1D case, and we verified numerically that the agreement is excellent already for $L$ and $L_{\rm leads}$ of the order of a few tens of sites.
The case of a simple square potential, where $V=V_0$ only on a region of size $L \times L$ and $V=0$ elsewhere, is less straightforward and attempting an analytical solution is beyond the scope of the present work. However, one would expect -- and indeed numerical simulations suggest -- that the quasiperiodic behaviour of the number of localised states survives in that case too; the numerical results are unfortunately too limited in accessible system sizes to be conclusive on their own. 


On a more speculative note, one may wonder what happens when several FM insertions of different lengths are present in the AFM Ising chain that determines the on-site potential in the Falicov-Kimball-inspired model presented in the main text. Based on our results, one would expect the system to exhibit a quasiperiodically distributed number of quasilocalised states that weakly interact with one another via the exponentially decaying tails of their wave functions. The effects of such states on transport as well as on the many-body properties of the system are interesting directions for future work. 
%
%

\section*{Acknowledgements} 
We would like to thank Olav F. Sylju{\aa}sen and Attila Szabó for helpful discussions, and John Chalker for suggesting the connection with Levinson's theorem. C.G. was supported by the Aker Scholarship. A.\v{S}. acknowledges financial support from the Swiss National Science Foundation (Grant No.~199969). This work was supported in part by the Engineering and Physical Sciences Research Council (EPSRC) grants No.~EP/P034616/1, No.~EP/T028580/1 and No.~EP/V062654/1. 
%
%

\bibliography{quasiperiodic_arXiv.bbl}
%
%

\appendix


\section{\label{app:Analytical N_L}
Analytical derivation of $N_\mathrm{loc}$\\(direct calculation)}

We study an infinite tight-binding chain with constant hopping amplitude $t=1$ and on-site potential
\begin{equation}
V_n = \begin{cases}
     0, &\quad n \leq 0\\
      V_0, &\quad 1 \leq n \leq L\\
       0, &\quad n \geq L+1,
     \end{cases}
\end{equation}
where $n$ labels the sites and goes from $-\infty$ to $\infty$.

By solving the Schrödinger equation analytically, we derive an expression for the number of states localised on the gated segment. There are two bulk equations, one for the leads and one for the gated segment, given by
\begin{eqnarray}
- \psi_{n-1} -  \psi_{n+1} &=& E \psi_n,
\\
- \psi_{n-1} + V_0 \psi_n - \psi_{n+1} &=& E \psi_n.
\end{eqnarray}
Using the ansatz $\psi_n = Ae^{ikn}$ separately for the leads and the gated segment, we obtain the dispersions
\begin{align}
\begin{split}
    E_{\mathrm{gated}}(k) &= V_0 - 2\cos k,\\
    E_{\mathrm{leads}}(k) &= -2\cos k
\, .
    \label{eq:dispersion}
\end{split}
\end{align}

Being interested in studying the states which localise on the gated segment and decay exponentially into the leads, we construct the following ansatz,
\begin{equation}\label{eq:WFansatz1}
\psi_n = \begin{cases}
      \hspace{8pt}A \, e^{-(p+i\pi)(L+1)/2}e^{(p+i\pi)n}, &\quad n \leq 0\\
       \hspace{8pt}B \nobarfrac{\cos}{\sin} \left[ kn - k (L+1)/2 \right], &\quad 1 \leq n \leq L\\
       \pm A \, e^{(p+i\pi)(L+1)/2}e^{-(p+i\pi)n}, &\quad n \geq L+1,
     \end{cases}
\end{equation}
where we have taken advantage of the fact that the potential is symmetric about the middle of the gated region,  meaning that the eigenstates must be symmetric or antisymmetric about $(L+1)/2$. Note that $(L+1)/2$ is not restricted to integer values and the following derivation therefore holds irrespective of $L$ being even or odd. Note also that there are only \textit{two possible choices} for the oscillatory phase $\varphi$ of the wave function in the leads that give real energies: $0$ and $\pi$. In our case, all the possible localised states are captured by $\varphi = \pi$ and we will thus only consider this case here ($\varphi = 0$ would only give $E < -2$).  Having explicitly accounted for the oscillating part of the wave function in the leads, we consider only real $p > 0$. The state in Eq.~\eqref{eq:WFansatz1} has two unknowns: $k$ and the relation between $A$ and $B$ ($p$ is given by requiring $E_{\mathrm{leads}}(p) =  2\cosh p = E_{\mathrm{gated}}(k)$; namely, $p(k) = \arccosh \left[V_0/2 -  \cos k\right]$). To find these unknowns, we have to impose the boundary equations which couple the leads and the gated segment. These are in general four equations (two equations for each boundary), but since we have already imposed the exchange symmetry of the wave function, we only need to consider the equations for one of the boundaries. The boundary equations for the left boundary read
\begin{align}
-\psi_{-1} - \psi_{1} &= E \psi_0, \label{eq:boundary1gated}\\
-\psi_{0} + V_0 \psi_{1} - \psi_{2} &= E \psi_{1}
\, . \label{eq:boundary2gated}
\end{align}

Inserting the ansatz Eq.~\eqref{eq:WFansatz1} into Eqs.~\eqref{eq:boundary1gated} and \eqref{eq:boundary2gated}, we arrive at the following equation for $k$:
\begin{equation}\label{eq:k-eq}
    -e^{p(k)} = \frac{ \nobarfrac{\cos}{\sin}\left[k - k\frac{L+1}{2} \right]}{ \nobarfrac{\cos}{\sin} \left[ - k\frac{L+1}{2} \right]},
\end{equation}
with $p(k) = \arccosh \left[V_0/2 -  \cos k\right]$.

This equation in general has several discrete solutions $k_i$, and each $k_i$ corresponds to a state localised on the gated segment as long as $p(k_i) > 0$. Finding an explicit expression for the solutions $k_i$ is difficult, but we can count the number of solutions by considering the behaviour of the functions on the two sides of the equation. 

\begin{figure}
\centering
\includegraphics[width=\columnwidth]{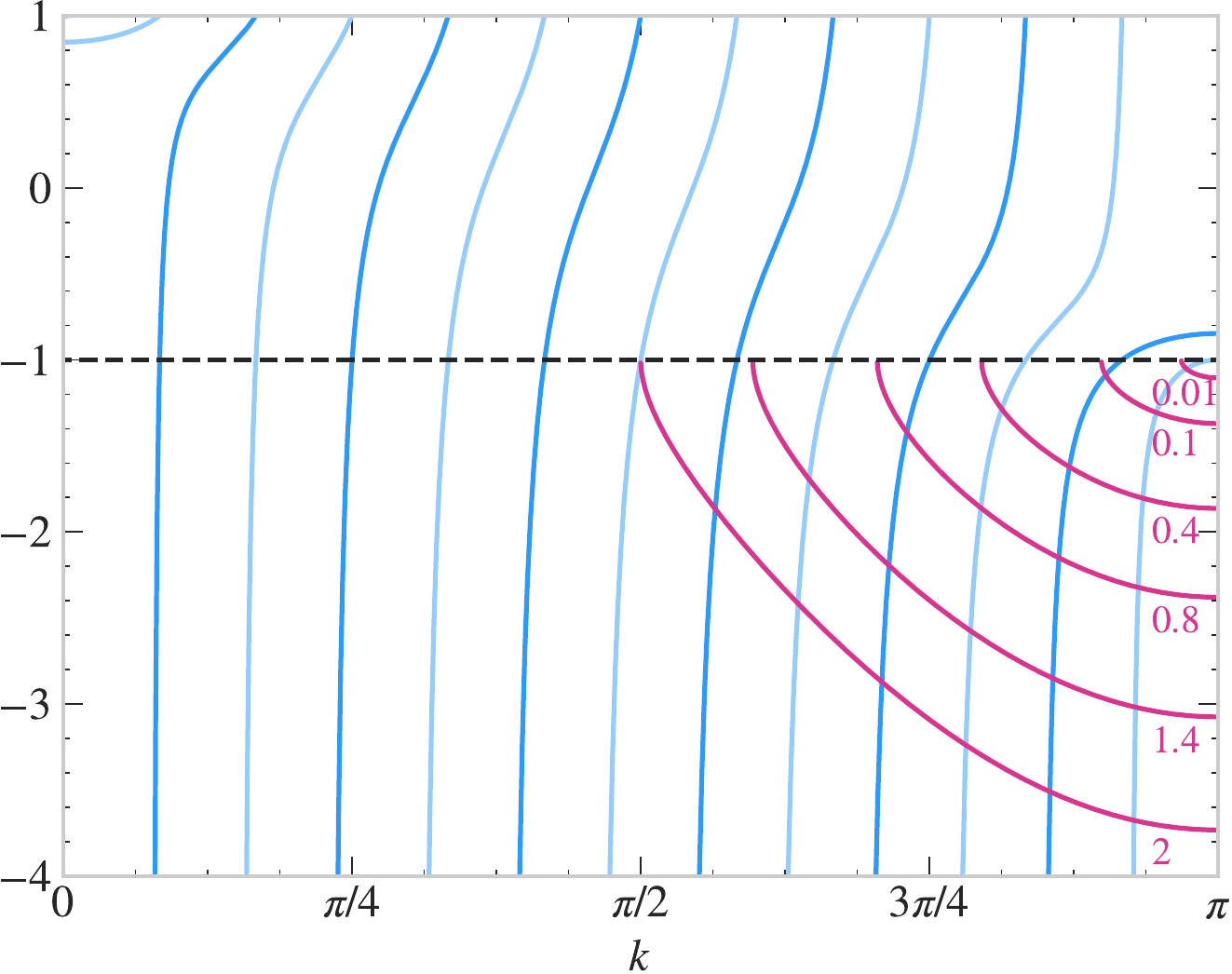}
\caption{The pink lines illustrate $\mathrm{LHS}(k)$ for different $V_0$.  The blue lines represent $\mathrm{RHS}(k)$ for $L = 12$. The black dashed line shows $C(k) = -1$. The blue lines (RHS) cross this line for $q_n = \pi/L n$. The pink lines touch $C(k)$ for $\alpha\pi$.}
\label{fig:solve_k_equation}
\end{figure}

The left-hand side 
\begin{equation}
\mathrm{LHS}(k) = -e^{\arccosh \left[V_0/2 -  \cos k\right]}
\end{equation}
is a function starting at $\mathrm{LHS}(k)=-1$, for $k = \arccos{\left(V_0/2 - 1\right)} \equiv \alpha \pi$, and decreasing monotonically as $k\to \pi$; see Fig.~\ref{fig:solve_k_equation}. For convenience, we define the starting point of $\mathrm{LHS}$ as a constant function 
\begin{equation}
    C(k) = -1.
\end{equation}

The right-hand side (of which we have one for the symmetric eigenstates and another for the antisymmetric eigenstates)
\begin{equation}
\mathrm{RHS}(k) = \frac{ \nobarfrac{\cos}{\sin}\left[k - k\frac{L+1}{2} \right]}{ \nobarfrac{\cos}{\sin} \left[ - k\frac{L+1}{2} \right]}
\end{equation}
is a function with derivative $\mathrm{RHS}'(k) \geq 0$ for $k \geq 0$ (and $\mathrm{RHS}'(k)=0$ only for $k = 0, \pi$). 

The functions $\mathrm{LHS}(k)$ and $\mathrm{RHS}(k)$ are illustrated in Fig.~\ref{fig:solve_k_equation}, where we plot $\mathrm{RHS}(k)$, which is independent of $V_0$, for $L = 12$, together with $\mathrm{LHS}(k)$ for a range of different values of $V_0$. 

To find an analytical expression for the number of solutions $k_i$ for an arbitrary length $L$ of the gated segment, let us start from the limit $V_0 = 0$ and increase $V_0$ continuously to see how the number of solutions increases. For $V_0 = 0$ there are no solutions ($\alpha = 1$ and $k=\pi=\alpha\pi$ solves the equation, but this solution is delocalised as $p(\alpha\pi) = 0$ with $E_{\mathrm{gated}}(\alpha\pi) = 2$). As one then lets $V_0 > 0$, one immediately finds an allowed solution for the $\mathrm{RHS}$ line crossing $C(k)$ at $\pi$, but for a $k_i$ slightly smaller than $\pi$ ($k_i$ decreases continuously from $\pi$ as $V_0$ increases, approaching the divergence of the corresponding RHS line; see Fig.~\ref{fig:solve_k_equation}). We then have only one solution until $V_0$ has increased enough for the next RHS line to fulfil 
\begin{equation}\label{eq:requirement}
C(\alpha \pi) = \mathrm{RHS}(\alpha\pi)
\, .
\end{equation}
In fact we first get a new allowed solution when $V_0$ is infinitesimally larger than the value solving Eq.~\eqref{eq:requirement}, as $k = \alpha\pi$ gives $p(\alpha\pi) = 0$.  Then, upon further increasing $V_0$, the number of solutions stays constant (but the values of $k_i$ change continuously) until $V_0$ is larger than the next value solving  Eq.~\eqref{eq:requirement}. Consequently,  the number of solutions of Eq.~\eqref{eq:k-eq} is the number of solutions $C(q) = \mathrm{RHS}(q)$ with $q > \alpha \pi$.  This follows from the fact that the derivative of RHS is positive, as illustrated in Fig.~\ref{fig:solve_k_equation}. Here, we see that once an RHS line crosses $-1$ for $q>\alpha\pi$, it must necessarily also cross $\mathrm{LHS}(k)$ for some $k = k_i$ smaller than $q$ and it must give us an allowed solution for this $k_i$. 

It can easily be shown that $C(q) = \mathrm{RHS}(q)$ for $q_n = \pi n / L$ with $n \in \{-L, ..., -1, 1, ..., L\}$. Counting the number of solutions now reduces to counting how many $q_n = \pi n / L > \alpha \pi$. This set of $q$'s is indeed not the values $k_i$ solving Eq.~\eqref{eq:k-eq}, and thus not the proper $k$'s characterising the localised eigenstates on the gated segment; they are simply the values which solve $C(q) = \mathrm{RHS}(q)$.  No physical significance should be given to these $q$'s other than them being a tool to count the number of solutions of Eq.~\eqref{eq:k-eq}. Negative $k$ gives the same physical state as positive $k$ and Eq.~\eqref{eq:k-eq} is symmetric for $k\to -k$; thus we only need to count the positive solutions. In summary, we arrive at the condition for solutions of the Schr\"{o}dinger equation localised on the gated segment 
\begin{equation}
    n > \alpha L.
\end{equation}
As $0 < n \leq L$, the number of such solutions is given by 
\begin{equation}
N_\mathrm{loc} = L - \mathrm{floor}(\alpha L).
\end{equation}
%
%

\section{\label{app:Analytical N_L Levinson}
Analytical derivation of $N_\mathrm{loc}$\\(Levinson's theorem)}

Levinson's theorem relates the difference in the phase shift of the extended states at zero and infinite momentum to the number of bound states. We here show that Levinson's theorem, appropriately modified for our lattice model, can be used to obtain the number of localised states, and it reproduces our earlier result, Eq.~\eqref{eq:exact_solution}. To show this, let us consider the same model as in App.~\ref{app:Analytical N_L}, with finite size for convenience, and focus on the states that are extended on the leads. We write one ansatz for the antisymmetric states ($\psi^A$) and one for the symmetric states ($\psi^S$):
\begin{equation}\label{eq:ansatzA}
\psi^{A}_n = \begin{cases}
     A\sin\left[
       kn - k\frac{L+1}{2} + \delta^A(k) 
       \right], & -L_\mathrm{leads} < n \leq 0\\
       B\sin\left[k^\prime n - k^\prime\frac{L+1}{2} \right], & 1 \leq n \leq L,
     \end{cases}
\end{equation}
\begin{equation}\label{eq:ansatzS}
\psi^S_n = \begin{cases}
     A\cos\left[kn - k\frac{L+1}{2} + \delta^S(k) \right], & -L_\mathrm{leads} < n \leq 0\\
     B\cos\left[k^\prime n - k^\prime\frac{L+1}{2} \right], & 1 \leq n \leq L,
     \end{cases}
\end{equation}
where the form of the wave function on the right lead follows by symmetry. $\delta^{A(S)}(k)$ is the scattering phase shift defined so that it vanishes for $V_0 = 0$. $k^\prime$ is given by $k^\prime = \arccos(V_0/2 + \cos k)$ and is imaginary for $-2 \leq E < -2+V_0$ and real for $-2+V_0 \leq E \leq 2$.

The open boundary conditions of the chain are imposed by assuming $\psi(-L_\mathrm{leads}) = \psi(L_\mathrm{leads}) = 0$. Consequently, the allowed values of $k$ are those where either of the functions
\begin{align}
f^A(k) &= \left(-kL_\mathrm{leads} - k\frac{L+1}{2} + \delta^A(k)\right)/\pi,\label{eq:fA} \\
f^S(k) &= \left(-kL_\mathrm{leads} - k\frac{L+1}{2} + \delta^S(k)\right)/\pi -\frac{1}{2}, \label{eq:fS}
\end{align}
take an integer value.

For sufficiently large $L_\mathrm{leads}$, $f^{A(S)}(k)$ is a monotonically decreasing function of $k$ (we require $\frac{\partial \delta}{\partial k} < L_\mathrm{leads} + \frac{L+1}{2}$, which includes the requirement that $\delta(k)$ is continuous). 
If we consider $f^{A(S)}(k)$ at the minimum and maximum momentum, i.e., $k = 0$ and $k = \pi$, we can then count the number of antisymmetric (symmetric) extended states as
\begin{equation}\label{eq:Nextmomentum}
N^{A(S)}_\mathrm{ext} = \mathrm{ceil} [f^{A(S)}(k=0)] - \mathrm{floor} [f^{A(S)}(k=\pi)] - 1.
\end{equation}
We have here paid extra caution to states with momenta $k = 0$ and $k=\pi$, as necessary. Their wave functions are zero everywhere and should thus not be included in the counting. This explains the choices of $\mathrm{ceil}$ and $\mathrm{floor}$ and why we subtract one in $N^{A(S)}_\mathrm{ext}$. 

By imposing the boundary equations between the leads and the gated segment, we find that the phase shift is given by
\begin{align}
\cot\left[\delta^A(k) - k\frac{L+1}{2}\right] &= \frac{\sin[k^\prime - k^\prime(L+1)/2]}{\sin[- k^\prime(L+1)/2]\sin k} - \cot k, \label{eq:cotdeltaA}\\
\tan\left[\delta^S(k) - k\frac{L+1}{2}\right] &= -\frac{\cos[k^\prime - k^\prime(L+1)/2]}{\cos[- k^\prime(L+1)/2]\sin k} + \cot k, \label{eq:tandeltaS}
\end{align}
where an appropriate integer number of $\pi$ should be added to the phase shift so that it is continuous.

For $V_0 = 0$, the phase shift is necessarily zero and all states are extended. We then have 
\begin{align*}
\left.N^A_\mathrm{ext}\right|_{V_0 = 0} &= L_\mathrm{leads}-1+\mathrm{ceil}[(L+1)/2],\\
\left.N^S_\mathrm{ext}\right|_{V_0 = 0} &= L_\mathrm{leads}+\mathrm{ceil}[L/2],
\end{align*}
giving $2L_\mathrm{leads} + L$ states in total.
As the total number of antisymmetric and symmetric states both should be independent of $V_0$, we find that the number of localised states is
\begin{align}
N^A_\mathrm{loc} = &  \begin{cases}
     \frac{1}{\pi}\left[\delta^A(k = \pi) - \delta^A(k = 0)\right] + \frac{1}{2} , &\quad L \; \mathrm{even}\\
      \frac{1}{\pi}\left[\delta^A(k = \pi) - \delta^A(k = 0)\right] , &\quad L \; \mathrm{odd},
     \end{cases} \label{eq:NlocAgated}\\
N^S_\mathrm{loc} = &  \begin{cases}
     \frac{1}{\pi}\left[\delta^S(k = \pi) - \delta^S(k = 0)\right] + \frac{1}{2} , &\quad L \; \mathrm{even}\\
      \frac{1}{\pi}\left[\delta^S(k = \pi) - \delta^S(k = 0)\right] + 1 , &\quad L \; \mathrm{odd},
     \end{cases} \label{eq:NlocSgated}
\end{align}

\begin{figure}
    \centering
    \includegraphics[width=\columnwidth]{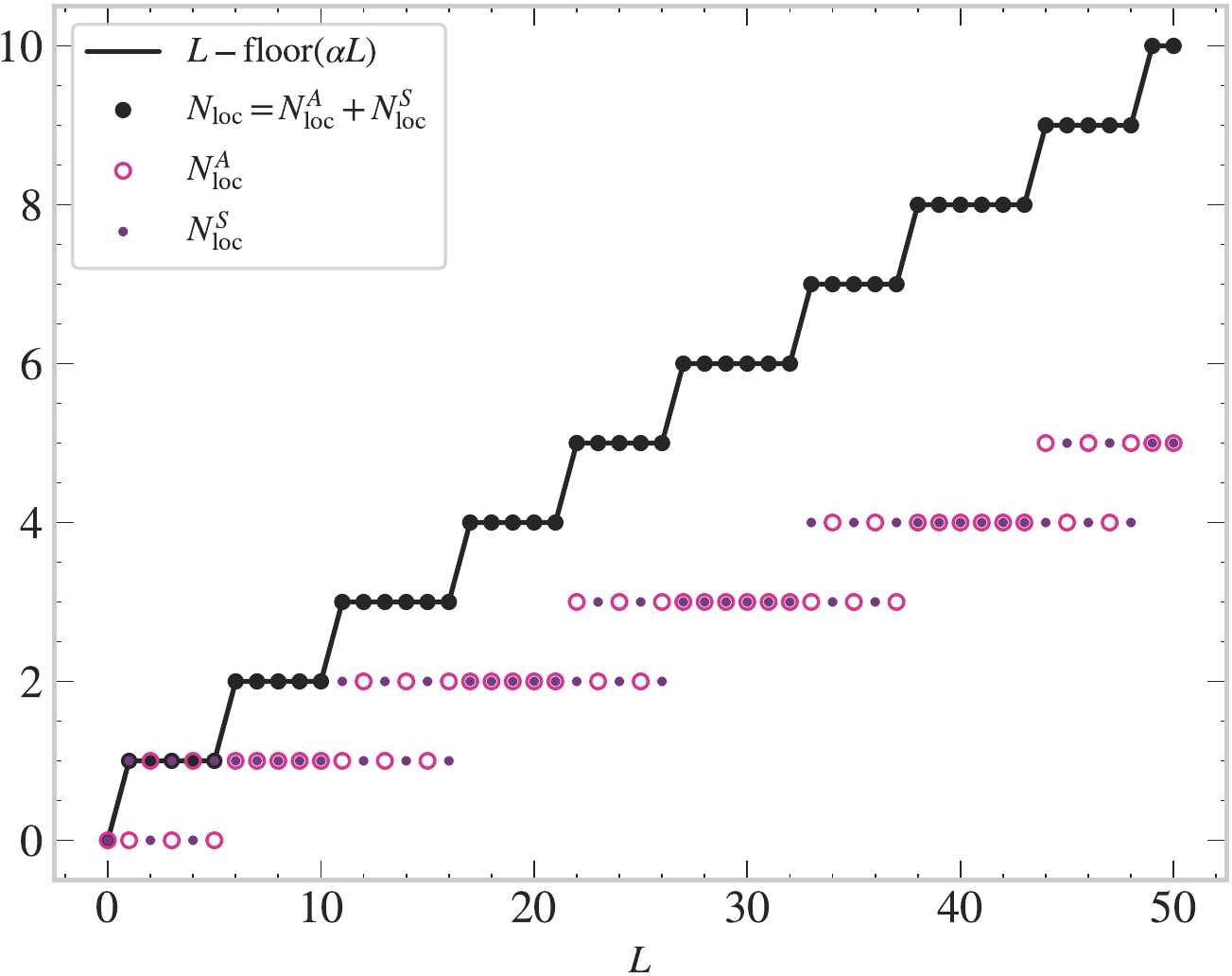}
    \caption{The number of localised states found from Levinson's theorem compared to $L- \mathrm{floor}(\alpha L)$ for $V_0 = 0.33$.}
    \label{fig:Levinson033}
\end{figure}
\begin{figure}
    \centering
    \includegraphics[width=\columnwidth]{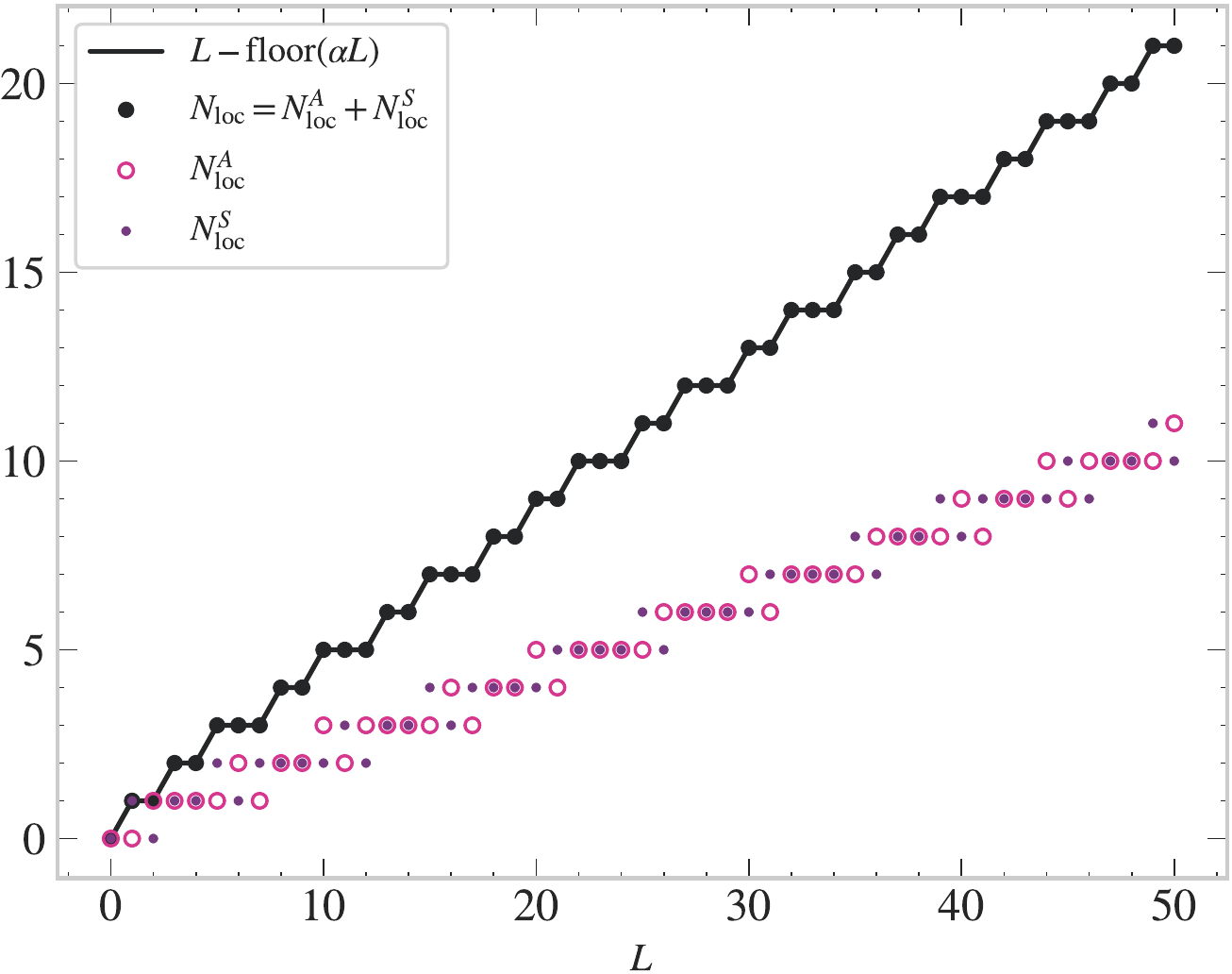}
    \caption{The number of localised states found from Levinson's theorem compared to $L- \mathrm{floor}(\alpha L)$ for $V_0 = 1.45$.}
    \label{fig:Levinson145}
\end{figure}

The total number of localised states is $N_\mathrm{loc} = N^A_\mathrm{loc} + N^S_\mathrm{loc}$. Numerically solving for $\delta(k)$, we end up with the number of localised states shown in Figs.~\ref{fig:Levinson033} and \ref{fig:Levinson145}, which is in perfect agreement with Eq.~\eqref{eq:exact_solution}.
%
%

\section{\label{app:DependenceOnV0} 
Dependence on $V_0$}
\begin{figure}
    \centering
    \includegraphics[width=\columnwidth]{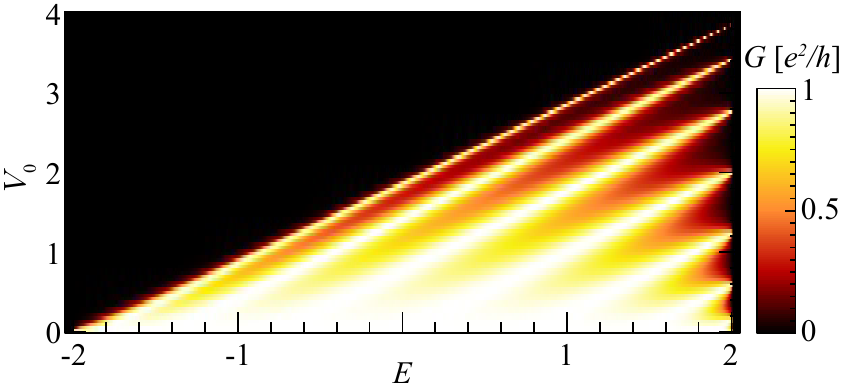}
    \caption{Conductance as a function of $V_0$ for $L = 10$, in the gated chain model in Sec.~\ref{sec:gated_segment}.}
    \label{fig:conductancs_vs_V0}
\end{figure}

Figure~\ref{fig:conductancs_vs_V0} shows the dependence of peaks in the conductance, from which the number of localised states can be extracted, on $V_0$. Notice that the behaviour of the system depends continuously on the gated voltage, and $N_\mathrm{loc}(V_0)$ does not show quasiperiodic behaviour as a function of the gate voltage $V_0$. 
%
%

%
%

\section{\label{app:Analytical cond}
Analytical calculation of the conductance}

The conductance through a single-channel 1D system is given by the Landauer formula 
\begin{equation}
    G(E) = \frac{e^2}{h} \mathcal{T}(E) \mathcal{T}^{*}(E)
\, , 
\end{equation}
where $E$ is the energy of the incident particle and $\mathcal{T}$ is the transmission coefficient through the gated region. To obtain the transmission coefficient, we first write the wave function of the free particle in continuum in the left and right leads (L, R) and in the gated region (M)
\begin{align}
    \psi_\mathrm{L}(x) &= e^{i k x} + \mathcal{R} e^{- i k x} \\
    \psi_\mathrm{M}(x) &= A e^{i k' x} + B e^{- i k' x} \\
    \psi_\mathrm{R}(x) &= \mathcal{T} e^{i k x} \, ,
\end{align}
where $k$ and $k'$ are the momenta of the leads and the middle region, respectively. Using the following boundary conditions, 
\begin{align}
\begin{split}
    \psi_\mathrm{L}(0) = \psi_\mathrm{M}(0);&  \qquad  \psi_\mathrm{M}(L) = \psi_\mathrm{R}(L) \\
    \partial_x \psi_\mathrm{L}(0) = \partial_x \psi_\mathrm{M}(0);&  \qquad  \partial_x \psi_\mathrm{M}(L) = \partial_x \psi_\mathrm{R}(L) \, ,  
\end{split}
\end{align}
we obtain the expression for the transmission coefficient 
\begin{equation}
    \mathcal{T} = \left[ 1 - \left( \frac{1+\kappa}{1-\kappa} \right)^2 \right] \frac{e^{-i k L}}{e^{i k'L} - \left( \frac{1+\kappa}{1-\kappa} \right)^2 e^{-i k' L}}\, ,
\end{equation}
where $\kappa \equiv k/k'$. The energy dependence is obtained by inverting Eqs.~\eqref{eq:dispersion}, which gives
\begin{align}
\begin{split}
    k(E) &= \arccos \left( -\frac{E}{2} \right) \\
    k'(E) &= \arccos \left( \frac{V_0-E}{2} \right) \\
    \kappa(E) &= \frac{\arccos \left( -\frac{E}{2} \right)}{\arccos \left( \frac{V_0-E}{2} \right)} \, .
\end{split}
\end{align}
The energy-dependent conductance then follows,
\begin{align}
    G(E) =  
    \frac{\frac{e^2}{h}   \left[ 1 - \left( \frac{1+\kappa}{1-\kappa} \right)^2 \right]^2}{1 + \left( \frac{1+\kappa}{1-\kappa} \right)^4 - 2 \left( \frac{1+\kappa}{1-\kappa} \right)^2 \cos\left[ 2 \arccos\left( \frac{V_0-E}{2} \right) L \right]} \, .
\end{align}
Note that $G(E)$ is an oscillatory function with maxima reaching $\frac{e^2}{h}$ located at energies $E_n$ that maximise the cosine term in the denominator, namely when 
\begin{equation}
    \arccos \left( \frac{V_0-E_n}{2} \right) = \pi \frac{n}{L} \, ,
\end{equation}
with $n \in \mathbb{N}$.
To count the total number of oscillations in the conductance for a given length $L$, we first need to restrict the energies $E_n$ to an interval $[-2, 2]$, which is the bandwidth of the leads. The number of oscillations in conductance for a fixed length $L$ is 
\begin{eqnarray}
    N_{\rm peaks}(L) &=& \left\lfloor \frac{L}{\pi}  \arccos \left( \frac{V_0}{2} - 1 \right) \right\rfloor 
\\ 
&=& \mathrm{floor}\left( \alpha L \right) 
    \, , 
    \label{eq:N_peaks}
\end{eqnarray}
with the same definition of $\alpha$ as in the main text and in App.~\ref{app:Analytical N_L}.
%
%

\section{\label{app:Analytical N_L^FM}
Analytical derivation of $N^{\mathrm{FM}}_\mathrm{loc}$}

The second example we consider in the main text is an infinite tight-binding chain with constant hopping amplitude $t=1$ and on-site potential
\begin{equation}
V_i = \begin{cases}
     (-1)^{i+1}W, &\quad i \leq 0\\
      +W &\quad 1 \leq i \leq L\\
     (-1)^{i+L+1}W , &\quad i \geq L+1,
     \end{cases}
\,
\label{eq:FMpotential}
\end{equation}
where $i$ labels the sites and goes from $-\infty$ to $\infty$. This can be viewed as a hopping particle interacting with classical Ising spins living on the sites of the chain, arranged in an infinite antiferromagnetic (AFM) pattern with a ferromagnetic (FM) insertion of length $L$. (Notice that, for convenience, we adopted a labelling convention for the potential in Eq.~\eqref{eq:FMpotential} such that $L=0$ corresponds to an infinite AFM pattern.)

We analytically derive the expression for the number of states localised on the FM segment by solving the Schr\"{o}dinger equation. In this case, the potential is symmetric about $i = (L+2)/2$. Thus, we only need to solve for $i \leq (L+2)/2$ and enforce a symmetric/antisymmetric wave function. Note that $(L+2)/2$ is not restricted to integer values and the following derivation therefore holds irrespective of $L$ being even or odd.

The AFM leads are divided into two sublattices. We therefore have three bulk equations: Two for the leads and one for the FM segment, given by
\begin{align}
- \psi^{-}_{n-1} + (+W) \psi^{+}_n - \psi^{-}_{n} &= E \psi^{+}_n,\\
-\psi^{+}_{n} + (-W) \psi^{-}_n - \psi^{+}_{n+1} &= E \psi^{-}_n,\\
-\psi^{FM}_{n-1} + (+W) \psi^{FM}_{n} - \psi^{FM}_{n+1} &= E \psi^{FM}_{n}
\, ,
\end{align}
where $\psi^{+}$ ($\psi^{-}$) is the wave function on the sublattice of potential $+W$ ($-W$) in the left AFM lead and $\psi^{FM}$ is the wave function in the FM segment. We have introduced a new index $n$ to label the unit cells, implying $i = 2n - 1, 2n$ in the left lead whereas $i = n$ in the FM segment. 

Using the ansatz $\psi^{+}_n = Ae^{ikn}$ for the leads, we get the dispersion
\begin{equation}
E_{\mathrm{AFM}}(k) = \pm \sqrt{W^2 + 2\left( 1 + \cos k \right)} \, , 
\end{equation}
and using the ansatz $\psi^{\mathrm{FM}}_n = Ae^{ikn}$ for the FM segment, we get the dispersion
\begin{equation}
E_{\mathrm{FM}}(k) = W - 2\cos k
\, .
\end{equation}

We are interested in finding the states which localise on the FM segment and therefore decay exponentially into the leads. To search for such solutions, we construct the following ansatz
\begin{align}
\psi^{+}_n &= Ae^{-(p+i\varphi)(L+2)/2}e^{(p+i\varphi) n},  \quad n \leq 0 \label{eq:WFansatzAFMlead}\\
\psi^{\mathrm{FM}}_n &= B \nobarfrac{\cos}{\sin} \left[ kn - k\frac{L+2}{2}\right],  \quad 1 \leq n \leq L+1 \label{eq:WFansatzFMsegment}
\end{align}
where the wave function on the right lead ($n > L$) follows from symmetry considerations.

Inserting the ansatz into the bulk equations, we get
\begin{equation}
E_{\mathrm{AFM}}(p,\varphi) = \pm \sqrt{W^2 + 2\left( 1 + \cosh (p + i\varphi) \right)}
\end{equation}
and
\begin{equation}
\psi^{-}_{n} = -\frac{(1+e^{p+i\varphi})}{E_{\mathrm{AFM}}(p,\varphi)+W}\psi^{+}_n
\, .
\end{equation}
To find the allowed values for $k$ in Eq.~\eqref{eq:WFansatzFMsegment}, we need to impose the boundary equations which couple the leads and the FM segment ($p$ is given by requiring $E_{\mathrm{AFM}}(p,\varphi) = E_{\mathrm{FM}}(k)$; namely, $p(k) = \arccosh \left[2\cos^2 k -2W\cos k - 1\right]$). 
The boundary equations for the left boundary read
\begin{align}\label{eq:boundary}
\begin{split}
-\psi^{+}_{0} - W \psi^{-}_{0} - \psi^{\mathrm{FM}}_{1} &= E \psi^{-}_0
\, , \\
-\psi^{-}_{0} + W \psi^{\mathrm{FM}}_{1} - \psi^{\mathrm{FM}}_{2} &= E \psi^{\mathrm{FM}}_{1}
\, .
\end{split}
\end{align}
Here, $k$ and $-k$ correspond to the same physical state, and we will only consider $k \geq 0$. Note also that there are only \textit{two possible choices for the oscillatory phase} $\varphi$ of the wave function in the leads that give real energies: $0$ and $\pi$. Here we need to consider the two cases separately: An oscillatory phase $\varphi = 0$ gives energies above the upper band, while an oscillatory phase $\varphi = \pi$ gives energies in the gap. 

Let us first consider the states above the upper band, $\varphi = 0$. Inserting the ansatz, Eqs.~\eqref{eq:WFansatzAFMlead} and \eqref{eq:WFansatzFMsegment}, into the boundary equations Eq.~\eqref{eq:boundary}, we get the following equation for $k$:
\begin{equation}\label{eq:k-eq2}
    2\left(\cos k - W \right)\frac{e^{p(k)}}{1 + e^{p(k)}} = \frac{ \nobarfrac{\cos}{\sin}\left[k - k\frac{L+2}{2} \right]}{ \nobarfrac{\cos}{\sin} \left[ - k\frac{L+2}{2} \right]}
\, ,
\end{equation}
with $p(k) = \arccosh \left[2\cos^2 k -2W\cos k - 1\right]$.
%
Again, we can quite easily count the number of solutions by considering the behaviour of the functions on the two sides of the equation. 

\begin{figure}
\centering
\includegraphics[width=\columnwidth]{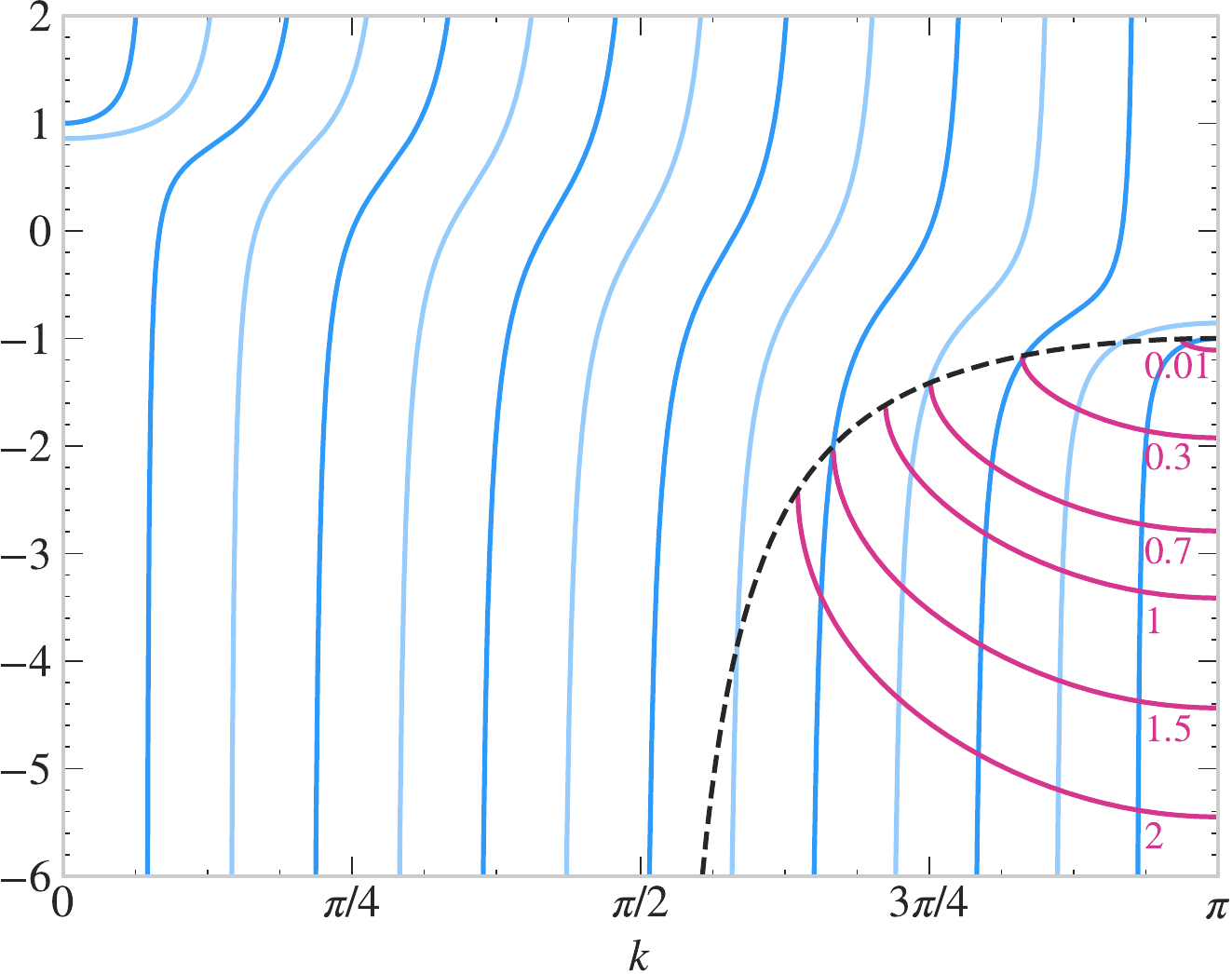}
\caption{The pink lines illustrate $\mathrm{LHS}^0(k)$ for different values of $W$. The blue lines represent $\mathrm{RHS}(k)$ for $L = 12$. The black dashed line shows $C(k)$. The blue lines (RHS) cross $C(k)$ for $q_n  \pi/L n$. The pink lines (LHS) touch $C(k)$ for $k = \beta\pi$.}
\label{fig:solve_k_equation0}
\end{figure}

The left-hand side 
\begin{equation}
\mathrm{LHS}^0(k) =  2\left(\cos k - W \right)\frac{e^{\arccosh \left[2\cos^2 k -2W\cos k - 1\right]}}{1 + e^{\arccosh \left[2\cos^2 k -2W\cos k - 1\right]}}
\end{equation}
is a function starting at $\mathrm{LHS}^0(k)=\cos k - W$ for $k = \arccos{\left[\left(-W-\sqrt{W^2 + 4}\right)/2\right]} \equiv \beta \pi$, which decreases monotonically as $k\to \pi$; see Fig.~\ref{fig:solve_k_equation0}. This is similar to what we had for the gated segment, but now the value of $\mathrm{LHS}^0(\beta\pi)$ varies with $W$. It can be shown that the line $C(k)$ giving the end point of LHS for a certain $W$ is given by
\begin{equation}
C(k) = \frac{1}{\cos k}
\, .
\end{equation}

The right-hand side (of which we have one for the symmetric eigenstates and another for the antisymmetric eigenstates)
\begin{equation}\label{eq:RHS}
\mathrm{RHS}(k) = \frac{ \nobarfrac{\cos}{\sin}\left[k - k\frac{L+2}{2} \right]}{ \nobarfrac{\cos}{\sin} \left[ - k\frac{L+2}{2} \right]}
\end{equation}
is a function with derivative $\mathrm{RHS}'(k) \geq 0$ for $k \geq 0$ (and $\mathrm{RHS}'(k)=0$ only for $k = 0, \pi$). 

The functions $\mathrm{LHS}^0(k)$ and $\mathrm{RHS}(k)$ are plotted in Fig.~\ref{fig:solve_k_equation0}, where we show $\mathrm{RHS}(k)$, which is independent of $W$, for $L = 12$, and $\mathrm{LHS}^0(k)$ for a range of different values of $W$. 

To find an analytical expression for the number of solutions $k_i$ of Eq.~\eqref{eq:k-eq2} for an arbitrary length $L$ of the FM segment, 
we need to find the solutions of 
\begin{equation}\label{eq:requirement2}
C(q) = \mathrm{RHS}(q)
\end{equation}
and count the number of solutions $q$ which are larger than $\beta\pi$. 
It is straight forward to show that $C(q) = \mathrm{RHS}(q)$ for $q_n = \pi n / L$ with $n \in \{\mathrm{ceil}(L/2), ..., L\}$.

Counting the number of solutions now reduces to counting how many $q_n = \pi n / L > \beta \pi$. Note again that we only allow solutions corresponding to real $p(k) > 0$, and $k = \beta\pi$ is thus not an allowed solution. We thus end up with
\begin{equation}
 N^0_\mathrm{loc} = L - \mathrm{floor}\left( \beta L \right)
\end{equation}
states localised on the FM segment with energies above the upper AFM band. 
 
\begin{figure}
\centering
\includegraphics[width=\columnwidth]{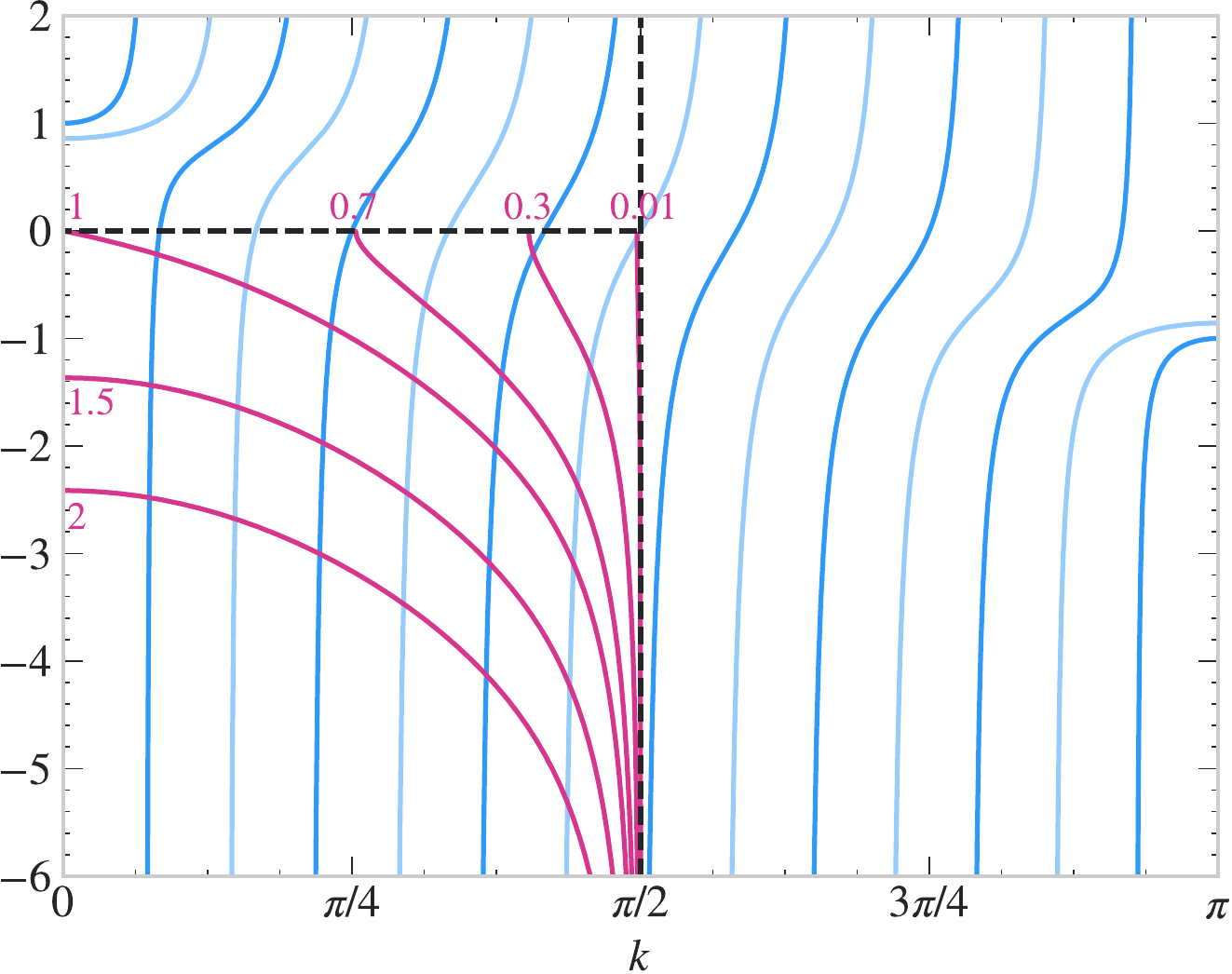}
\caption{The pink lines illustrate $\mathrm{LHS}^\pi(k)$ for different values of $W$. The blue lines represent $\mathrm{RHS}(k)$ for $L = 12$. The black dashed lines shows $C(k)=0$ and $k = \pi/2$. The blue lines (RHS) cross $C(k)$ for $q_n = \pi/L n$. The pink lines (LHS) touch $C(k)$ for $\gamma\pi$ and diverge to $-\infty$ for $k\to \pi/2$.}
\label{fig:solve_k_equationpi}
\end{figure}
 
To count the number of localised states with energies in the gap we must set $\varphi = \pi$. $\mathrm{LHS}(k)$ then becomes
\begin{equation}
\mathrm{LHS}^\pi(k) =  -2\left(\cos k - W \right)\frac{e^{\arccosh \left[-2\cos^2 k +2W\cos k + 1\right]}}{1 - e^{\arccosh \left[-2\cos^2 k +2W\cos k + 1\right]}}
\end{equation}
while $\mathrm{RHS}(k)$ stays the same as in Eq.~\eqref{eq:RHS}. $\mathrm{LHS}^\pi(k)$ starts at $k = \gamma\pi \equiv \arccos W$ with $\mathrm{LHS}^\pi(\gamma\pi) = 0$ ($k = 0$ for $W > 1$ with $\mathrm{LHS}^\pi(0) = -2(1-W)e^{\arccosh (2W-1)}/(1-e^{\arccosh (2W-1)})$) and decreases continuously as $k\to \pi/2$,  where it diverges to $-\infty$.

The functions $\mathrm{LHS}^\pi(k)$ and $\mathrm{RHS}(k)$ are plotted in Fig.~\ref{fig:solve_k_equationpi} for a range of different values of $W$. 

In this case, counting the number of solutions reduces to counting the number of solutions to $\mathrm{RHS}(q) = 0$ with $q > \gamma\pi$ which belongs to an RHS line that diverges to $-\infty$ for some $k < \pi/2$.  $\mathrm{RHS}(q) = 0$ for $q_n = \pi n / L$ with $n \in \{1, \dots, L-1\}$ and $\mathrm{RHS}(q^\infty) \to -\infty$ for  $q^\infty_n = \pi n / (L+2)$ with $n \in \{1, \dots,  L+1\}$. To find the number of solutions, we subtract the number of $q_n \leq \gamma\pi$ (for $W > 1$, we subtract the number of $q_n < 0$, which is trivially zero) from the number of $q^\infty_n < \pi/2$. Note that $k = \gamma \pi$ and $k = \pi/2$ are not allowed solutions, as these have $p = 0$. 


The total number of $q^\infty_n < \pi/2$ is $\mathrm{ceil}(L/2)$ and the total number of $q_n \leq \gamma\pi$ is $\mathrm{floor}\left(\gamma L\right)$.
We therefore have
\begin{equation}
N^{\pi}_\mathrm{loc} = \begin{cases}
    \mathrm{ceil}\left(L/2\right) - \mathrm{floor}\left(\gamma L\right), & W \leq 1\\
    \mathrm{ceil}\left(L/2\right), & W > 1
    \,
    \end{cases}
\end{equation}
states which localise in the gap.

The total number of states localised on the FM segment is thus
\begin{align}
    N^{\mathrm{FM}}_\mathrm{loc} =& N^{0}_\mathrm{loc} + N^{\pi}_\mathrm{loc} 
    \nonumber \\
    \!\!\!\! =& 
    \begin{cases}
    L - \mathrm{floor}\left(\beta L \right) + \mathrm{ceil}\left(L/2\right) - \mathrm{floor}\left(\gamma L\right), & W \leq 1\\
    L - \mathrm{floor}\left(\beta L \right) + \mathrm{ceil}\left(L/2\right), & W > 1
    \, .
    \end{cases}
\end{align}

%
%





%
%

\section{\label{app:WFProfiles} 
Wave function profiles}
We present here some illustrative examples of the eigenstate wave functions localised on the gated / ferromagnetic segment discussed in our work. 

\begin{figure}
    \centering
    \includegraphics[width=\columnwidth]{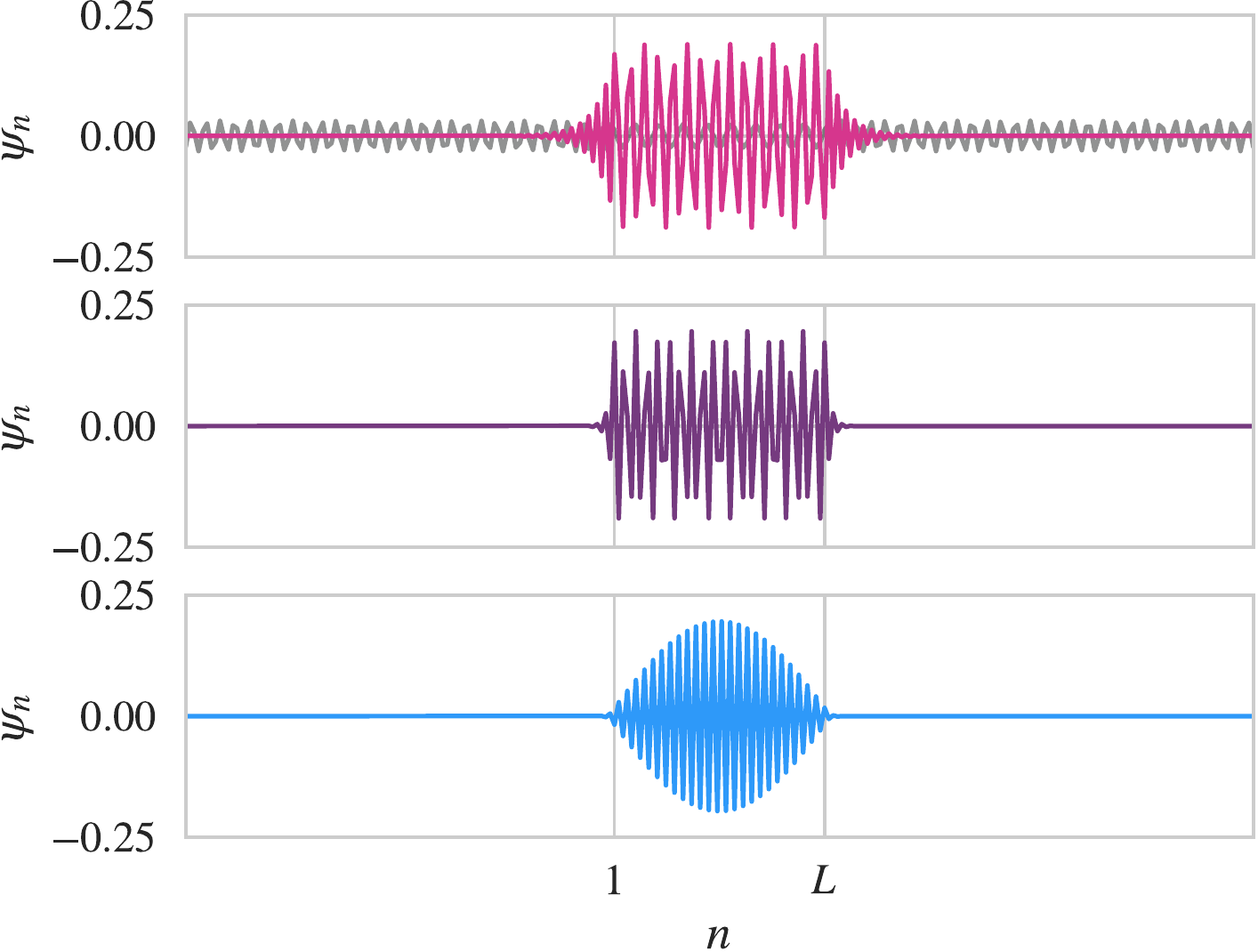}
    \caption{Examples of eigenstate wave functions localised on the gated segment for $V_0 = 1.45$ and $L = 50$. Top: $E = 2.06$. Middle: $E = 2.95$. Bottom: $E = 3.45$. The top panel also shows a delocalised state with $E = 0.566$ in light grey for comparison.}
    \label{fig:WFgated}
\end{figure}

Figure~\ref{fig:WFgated} shows examples of wave functions with energies $E > 2$ for the gated chain model in Sec.~\ref{sec:gated_segment}. The localised behaviour is clearly evident for $E>2$, when compared to a delocalised state for $E<2$ shown in the top panel for reference (light grey curve). 
%

\begin{figure}
    \centering
    \includegraphics[width=\columnwidth]{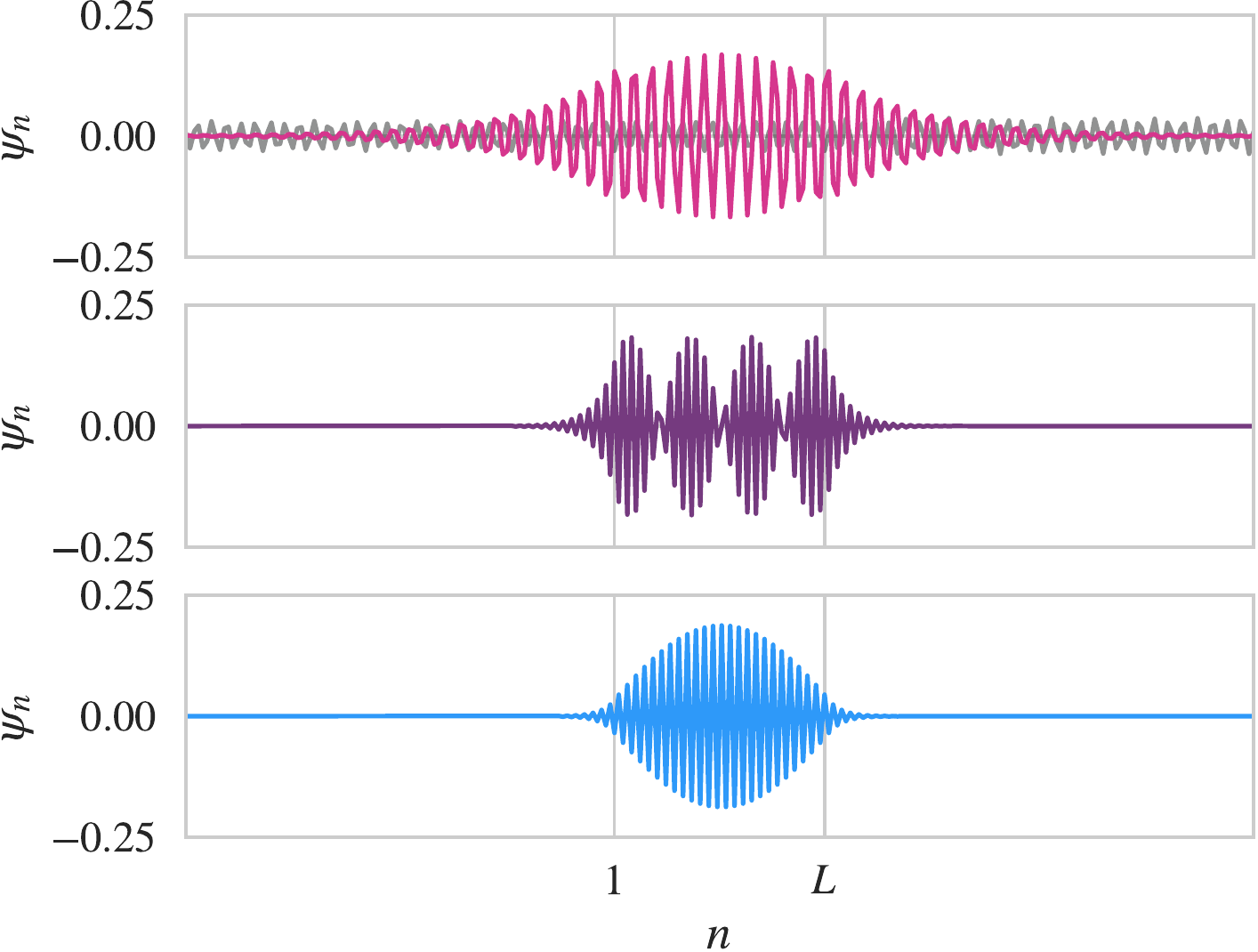}
    \caption{Examples of eigenstate wave functions localised on the ferromagnetic segment for $W = 0.1$ and $L = 50$. Top: $E = 0.0265$ (in the gap). Middle: $E = 2.05$. Bottom: $E = 2.10$. The top panel also shows a delocalised state with $E = 0.543$ in light grey for comparison.}
    \label{fig:WFFMinAFM}
\end{figure}

Figure~\ref{fig:WFFMinAFM} shows similar examples of eigenstate wave functions for the ferromagnetic segment in an antiferromagnetic chain model in Sec.~\ref{sec:antiferro_chain}. 
%

%
%

\section{\label{app:SpecialCases}
Special cases}

While the quasiperiodic behaviour discussed in the main text is generic, there are fine-tuned values of the system parameters when it disappears. For completeness, we review them here briefly for the two models considered in our work. 
%
%

\subsubsection{Gated segment}

In the case of the gated segment, $N_\mathrm{loc}$ shows \textit{periodic} fluctuations about its linear behaviour whenever $\alpha$ is rational, implying that $\alpha\pi = \arccos\left(V_0/2-1\right)$ is a rational number times $\pi$. This occurs for an infinite set of fine-tuned values of $V_0$ (e.g., $V_0 = \{2-\sqrt{2}, 1, 2\}$); however, it is a set of measure zero on the real line. 
%
%

\subsubsection{Ferromagnetic segment in an antiferromagnetic chain}

Let us now consider the example of a ferromagnetic segment in an antiferromagnetic chain. 
For $W \geq 1$, $N^{\mathrm{FM}}_\mathrm{loc}$ only depends on $\beta$ and the situation is very similar to that of the gated segment. If $\beta$ is rational, which happens for an infinite but sparse number of values $W$, $N^{\mathrm{FM}}_\mathrm{loc}$ shows periodic fluctuations. 
In the case of $W < 1$, $N^{\mathrm{FM}}_\mathrm{loc}$ depends on both $\beta$ and $\gamma$ and it only loses its quasiperiodic behaviour if both $\beta$ and $\gamma$ are rational. We only find one value of $W$ where this happens, namely $W = \sqrt{2} / 2$.

\vspace{3cm}

\end{document}